\documentclass[11pt,twoside]{article}

\usepackage{graphicx}
\usepackage{asp2006}
\usepackage{epsf}
\usepackage{psfig}
\usepackage{lscape}

\def \agile {AGILE}
\def \egret {EGRET}

\def \igr {INTEGRAL}
\def \swi {{\it Swift}}
\def \rem {REM}
\def \webt {WEBT}

\def \degmark{^\circ}

\def \phcmsec{\hbox{photons cm$^{-2}$ s$^{-1}$}}

\def \gray {$\gamma$-ray }
\def \source {\hbox{3C~454.3}}

\begin{document}
\title{Long-term \gray and multi wavelength observations of 3C~454.3
  (a.k.a. the {\it Crazy Diamond})}   %%% Fill in title
\author{S.\ Vercellone, on behalf of the AGILE Team}   %%% Fill in author names
\affil{INAF, Istituto di Astrofisica Spaziale e Fisica Cosmica, 
 Via U. La Malfa 153, I-90146 Palermo, Italy}    %%% Fill in author affiliations

\begin{abstract} %%% Abstract to run on from here.
During the period July 2007 - January 2009, the AGILE satellite,
together with several 
other space- and ground-based observatories monitored the activity of 
the flat-spectrum radio quasar 3C~454.3, yielding the longest 
multiwavelength coverage of this \gray quasar so far.
The source underwent an unprecedented period of very high activity above 
100 MeV, a few times reaching \gray flux levels on a day time scale 
higher than $F=400 \times 10^{-8}$\,\phcmsec, in conjunction with an extremely variable 
behavior in the optical $R$-band, even of the order of several tenth of 
magnitude in few hours, as shown by the GASP-WEBT light curves.
We present the results of this long term multiwavelength monitoring 
campaign, with particular emphasis on the study of possible lags between
the different wavebands, and the results of the modeling of simultaneous
spectral energy distributions at different levels of activity.
\end{abstract}

%%% MAIN BODY OF TEXT GOES HERE. CONSULT "INSTRUCTIONS FOR AUTHORS USING
%%% LATEX2E MARKUP", SECTIONS 2.3-2.6 FOR HELP WITH EQUATIONS, FIGURES,
%%% AND TABLES.

%%%%%%%%%%%%%%%%%%%%%%%%%%%
\section{Introduction}   %%% Top level section head (remove "%" symbol)
%%%%%%%%%%%%%%%%%%%%%%%%%%%
Among active galactic nuclei (AGNs), blazars show intense and variable
\gray emission above 100~MeV \citep{Hartman1999:3eg}, and the
variability time scale
can be as short as a few days, or last a few weeks.
They emit across several decades of energy, from the radio to the TeV
energy band and their spectral energy distributions (SEDs) are typically 
double humped with a first peak occurring in
the IR/optical band in the
flat-spectrum radio quasars (FSRQs) and low-energy peaked BL Lacs (LBLs),
and at UV/X-rays in the 
high-energy peaked BL Lacs (HBLs). This peak is commonly 
interpreted as synchrotron radiation
from high-energy electrons in a relativistic jet. The second SED
component is commonly interpreted as inverse Compton (IC) scattering 
of soft seed photons by relativistic electrons, and peaks in the
MeV--GeV and in the TeV energy bands in the FSRQs/LBLs and in the
HBLs, respectively. 
A recent review of the blazar emission mechanisms and energetics
is given in \cite{Celotti2008:blazar:jet}.

The FSRQ \source{} (PKS~2251$+$158; 
$z=0.859$) is certainly one of the most active extragalatic sources at
high energy. 
In the \egret{} era, it was detected in 1992
during an intense \gray flaring episode 
\citep{Hartman1992:3C454iauc, Hartman1993:3C454_EGRET}
when its flux $F_{\rm E>100MeV}$ was observed to vary within the range
$(0.4-1.4) \times 10^{-6}$\,photons\,cm$^{-2}$\,s$^{-1}$. In 1995, a 2-week 
campaign detected a \gray flux $< 1/5$ of its historical maximum
\citep{Aller1997:3C454_EGRET}.

\begin{figure}[!ht]
\begin{center}
%\vspace{-1truecm}
\includegraphics[width=8cm]{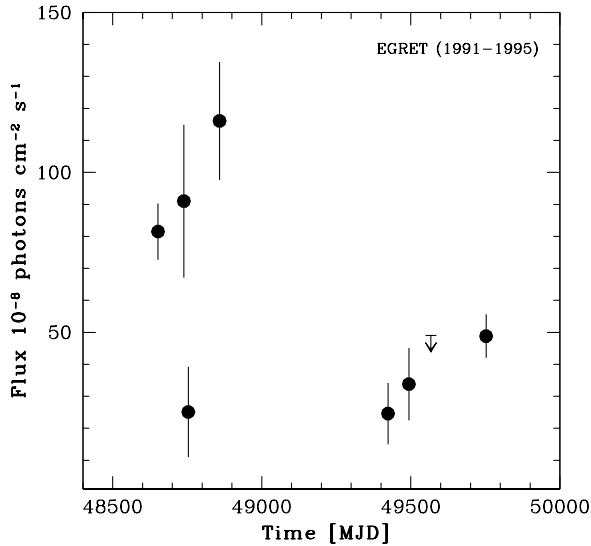}  
\end{center}
%\plotone{3c454_egret.eps}
\vspace{-0.5truecm}
\caption{EGRET \source{} light curve for $E>100$\,MeV in the period
  1991-1995. The downward arrow represents a $2\sigma$
      upper-limit. Data from \cite{Hartman1999:3eg}.} \label{Fig:3c454:egret}
\end{figure}

Figure~\ref{Fig:3c454:egret} shows the \gray light curve for
$E>100$\,MeV as observed by EGRET in the period 1991--1995.

In 2005, \source{} underwent a  major flaring activity in almost 
all energy bands (see \citealt{Giommi2006:3C454_Swift}).
In the optical, it reached $R=12.0$\,mag \citep{vil06} 
and it was detected by \igr{} at a flux\footnote{Assuming
a Crab-like spectrum.} level of
$\sim 3 \times 10^{-2}$\,photons\,cm$^{-2}$\,s$^{-1}$ in the 3--200~keV 
energy band \citep{Pian2006:3C454_Integral}. 
Since the detection of the exceptional 2005 outburst, several monitoring 
campaigns were carried out to follow the source multifrequency behavior 
\citep{vil06,vil07,rai07,rai08a,rai08b}. During the last of these campaigns, 
3C~454.3 underwent a new 
optical brightening in mid July 2007, which triggered observations at all 
frequencies, including the \agile{} one.
In the following, we briefly describe the \agile{} satellite, and we
discuss the various campaigns triggered on \source{}.

%

%%%%%%%%%%%%%%%%%%%%%%%%%%%
\section{The AGILE Satellite}   
%%%%%%%%%%%%%%%%%%%%%%%%%%%
The AGILE satellite \citep{Tavani2009:Missione}
is a mission of the Italian Space Agency (ASI) devoted to high-energy
astrophysics, operative since April 2007.
The scientific instrument combines four
active detectors yielding broad-band coverage from hard X-rays
to \gray:
a Silicon Tracker~\citep[ST;][30~MeV--30~GeV]{Prest2003:agile_st},
a co-aligned coded-mask hard X-ray imager, Super--AGILE
\citep[SA;][18--60~keV]{Feroci2007:agile_sa}, a non-imaging CsI
Mini--Calorimeter~\citep[MCAL;][0.3--100~MeV]{Labanti2009:agile_mcal},
and a segmented Anti-Coincidence System~\citep[ACS;][]{Perotti2006:agile_ac}.
The Gamma-Ray Imaging
Detector (GRID) reaches a 5-$\sigma$ sensitivity of about
$(1-2)\times10^{-6}$\,cm$^{-2}$\,s$^{-1}$ above 100~MeV for an
integration time of 2 days, sources positioned 30 degrees
off-axis, an Earth occultation fraction of 20\% and near the
Galactic Plane (high \gray background). The hard X-ray monitor on--board
AGILE, Super-\agile{} (SA), reaches a sensitivity of about 15 mCrab at 
5-$\sigma$ in the energy range 18--60~keV in a one-day observation.

%%%%%%%%%%%%%%%%%%%%%%%%%%%
\section{The AGILE Campaigns}   
%%%%%%%%%%%%%%%%%%%%%%%%%%%

%--------------------
\subsection{July 2007}   
%--------------------
At the epoch of the \source{} optical flare (2007 July 19--21) the
closest pointing position allowed by the solar panel
constraints ranged within 35$^{\circ}$--40$^{\circ}$ from the source.
Because of the remarkably large ($\sim 3$\,sr) field of view (FOV)
of the GRID and the successful
detection of the Vela pulsar at $\sim 55^{\circ}$ off-axis
obtained during the science verification phase, AGILE could
reliably study \source{} despite its large off-axis position.
The observations were performed between 2007 July 24 14:30 UT and 2007 
July 30 11:40 UT, for a total pointing duration of $\sim$220~ks.
The source detection significance is 
13.8$\sigma$ as derived from a maximum likelihood analysis
\citep{Vercellone2008:3C454_jul07}.

\begin{figure}[!ht]
%\plotone{3c454_lc_jul07.eps}
\begin{center}
%\vspace{-1truecm}
\includegraphics[width=9cm]{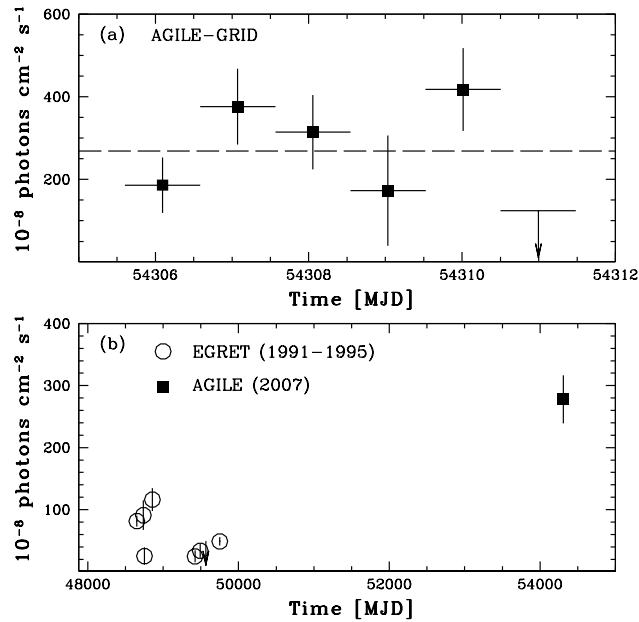}  
\end{center}
\vspace{-0.5truecm}
\caption{Panel (a): \source{} light curve in the period 2007 July
  24--30. Panel (b): comparison between the long term EGRET and AGILE
  average fluxes.
  Adapted from \cite{Vercellone2008:3C454_jul07}.} \label{Fig:3c454:lcjul07}
\end{figure}
Figure~\ref{Fig:3c454:lcjul07}, panel (a), shows the AGILE/GRID light
curve for $E>100$\,MeV during the \gray flare, while panel (b) shows
the comparison with the historical EGRET flux values.

We note that \source{} is detected at a 4-$\sigma$ level
during almost the whole period on a 1-day timescale. The 
average \gray flux above 100 MeV for the whole period is
$ F_{\rm E>100\,MeV} = (280 \pm 40) \times 10^{-8}$\,\phcmsec.
The average \gray flux above 100 MeV for the whole period was 
the highest ever detected from this
source, as shown in Figure~\ref{Fig:3c454:lcjul07}, panel (b).

The source was not detected (above 5-$\sigma$) by the Super-AGILE
Iterative Removal Of Sources (IROS) applied to the $Z$ image, in
the 20--60~keV energy range.
Assuming a power law spectral shape with photon index $\Gamma=1.5$, we
find a 3-$\sigma$ upper limit of $2.3 \times 10^{-3}$\,\phcmsec{}
on the average flux from \source{}.

%
%--------------------
\subsection{November 2007}   
%--------------------
%
In November 2007 \agile{} began pointing \source{} at high off-axis angle
(about $40\degmark$). Nevertheless, in a few days \source{} was 
detected at more than 5-$\sigma$,
exhibiting variable activity on a day time-scale.
Immediately after the source detection, a multiwavelength campaign
started \citep{Vercellone2009:3C454_nov07}.
\agile{} data were collected during two different periods, the
first ranging between  2007-11-10 12:17 UT and 2007-11-25 10:57 UT and
the second between 2007-11-28 12:05 UT and 2007-12-01 11:39 UT,
for a total of about 592\,ks.
The three-day gap between them was due to a pre-planned GRID 
calibration activity. 
\igr{} data were collected during a dedicated ToO on revolutions 
623 (between 2007-11-20 03:35 UT and 2007-11-22 08:46 UT) and 
624 (between 2007-11-22 20:45 UT and
2007-11-24 15:50 UT), for a total of about 300\,ks, while
\swi{}/XRT data were obtained during several ToO pointings for a total 
of about 10\,ks.
\webt{} data (radio to optical) as well as \swi/UVOT data were published 
in \cite{rai08a}, while
\rem{} data were acquired following a ToO request.
In both cases, optical data were acquired continuously during the whole
\agile{} campaign.
\begin{figure}[!ht]
%\plotone{3c454_lc_nov07.eps}
\begin{center}
%\vspace{-1truecm}
\includegraphics[width=9cm]{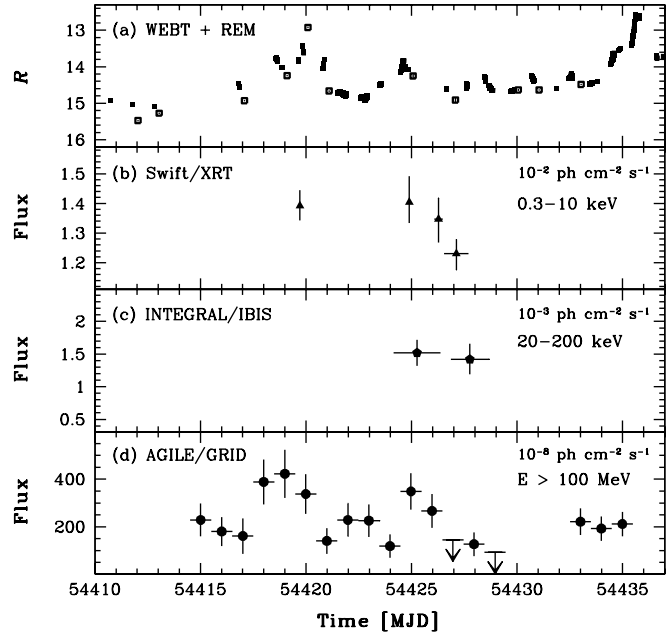}  
\end{center}
\vspace{-0.5truecm}
\caption{Simultaneous light curves acquired during the period
    2007 November 6--December 3. Circles represent \agile{}/GRID
    data ($E>100$~MeV); pentagons represent \igr{}/IBIS data
    (20--200~keV); triangles represent \swi{}/XRT data (0.3--10~keV);
    solid and open squares represent $R$-band WEBT \citep{rai08a} and REM
    data, respectively.
  Adapted from \cite{Vercellone2009:3C454_nov07}.} \label{Fig:3c454:lcnov07}
\end{figure}

Figure~\ref{Fig:3c454:lcnov07} shows the multi wavelength light curves
during the November 2007 observing campaign.
We investigated the expected $\gamma$-optical flux correlation by means of
the discrete correlation function (DCF).
The DCF peak occurred at $\tau=0$, and
its value is $\sim 0.5$. This indicates a moderate correlation, with no
significant time delay between the \gray and optical flux variations.

We fit the SEDs for two major \gray flaring episodes
by means of a one-zone leptonic model, considering the contributions 
from synchrotron self-Compton (SSC), and from external seed photons 
originating both from the accretion disk and from the broad-line
region (BLR).
We note that during both flaring episodes, the
external Compton scattering of direct disk radiation (ECD) 
contribution can account for the soft and hard X-ray portion of
the spectrum, which show a moderate, if any, time
variability. However, we note that the ECD component alone
cannot account for the hardness of the \gray spectrum.
We therefore  argue that in the \agile{} energy band a
dominant contribution from the external Compton scattering of
the BLR clouds (ECC) seems to provide a better fit of
the data during the \gray-ray flaring states.

%
%--------------------
\subsection{December 2007}   
%--------------------
%
In December 2007 \agile{} began a multi wavelength campaign involving 
{\it Spitzer}, {\it Swift}, {\it Suzaku}, the WEBT consortium, the REM and MITSuME 
telescopes, offering a broad band coverage that allowed a truly
simultaneous sampling of the synchrotron and IC
emissions \citep{Donnarumma2009:3C454_dec07}.

\begin{figure}[!ht]
%\plotone{3c454_sed_dec07.eps}
\begin{center}
%\vspace{-1truecm}
\includegraphics[width=9cm]{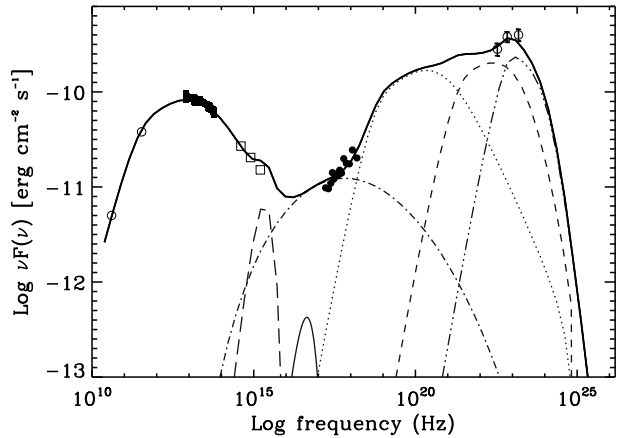}  
\end{center}
\vspace{-0.5truecm}
\caption{Simultaneous SED for 2007 December 13. 
  We note the possible contribution of the
  EC on the hot corona component (dash-dot-dot line) to the model fit.
  Adapted from \cite{Donnarumma2009:3C454_dec07}.} \label{Fig:3c454:seddec07}
\end{figure}

Figure~\ref{Fig:3c454:seddec07} shows the SED during 
2007 December 13, characterized by the broadest multi wavelength
coverage. During this epoch the source was in a \gray state
characterized by a flux for $E>100$~MeV of the order 
of $\sim 200 \times 10^{-8}$\,\phcmsec. 
We found that a model accounting for ECD and ECC components
does not fully describe the SEDs in the three epochs. An additional contribution, 
possibly from the hot corona with $T = 10^{6} K$ surrounding the jet, is required to 
account simultaneously for the synchrotron and the inverse Compton emissions 
during those epochs. 

Moreover, a detailed analysis of the correlation between the
flux variations in the \gray and optical energy band yields
a possible $\sim1$ day delay of the \gray emission with respect to 
the optical one. 

%
%--------------------
\subsection{Eighteen Months of Monitoring}   
%--------------------
%
Starting from May 2008, \agile{} conducted several multi wavelength
observing campaigns on \source{}. A detailed discussion of these
results is presented in \cite{Vercellone2009:3C454_18months}.

\begin{figure}[!ht]
\plotone{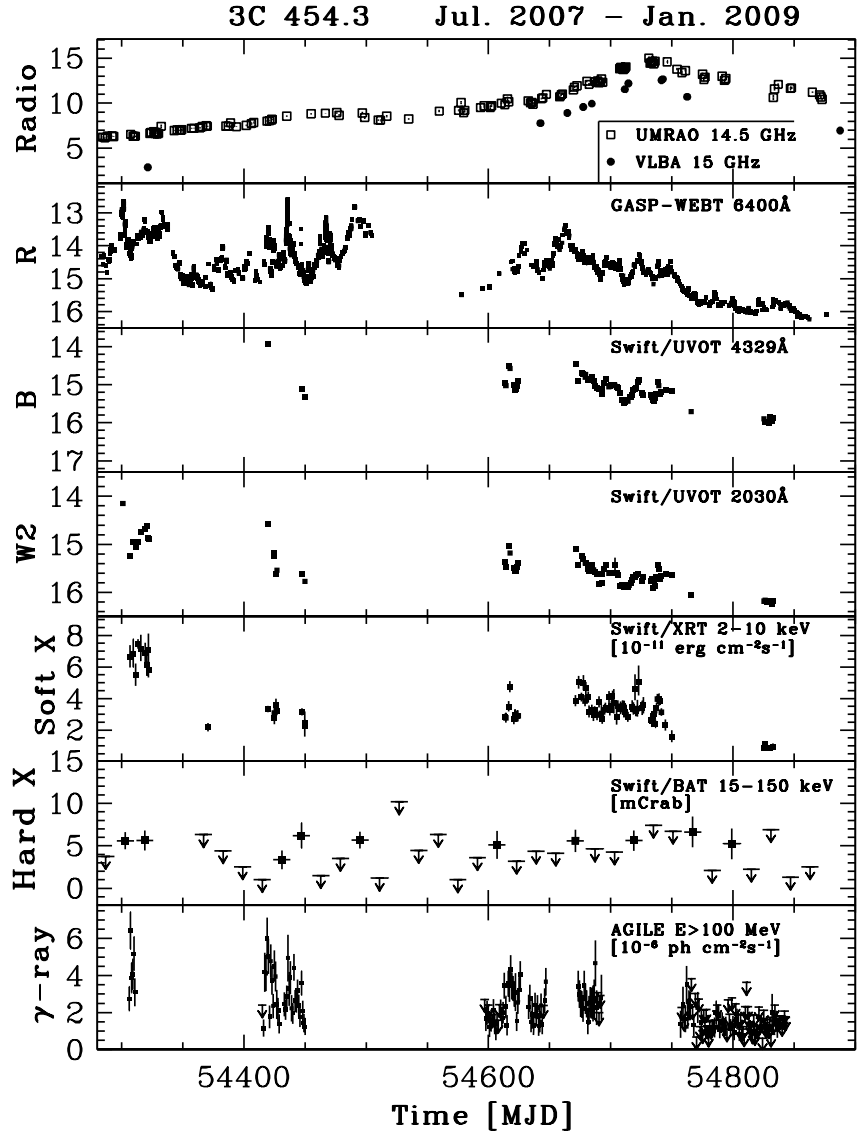}
\caption{Simultaneous SEDs for 2007 December 5 (light grey line), 12 (grey
  line), and 15 (black line). We note the possible contribution of the
  EC on the hot corona component (dash-dot-dot line) to the model fit.
  Adapted from \cite{Vercellone2009:3C454_18months}.} \label{Fig:3c454:18months}
\end{figure}

Figure~\ref{Fig:3c454:18months} shows the \source{} light curves 
at different energies over the whole period July 2007 -- January 2009.
The different panels
show, from bottom to top, the \agile{}/GRID light curve at
$\approx 1$-day resolution for E$>$100~MeV in units of
$10^{-8}$\,\phcmsec, the \swi{}/BAT light curve in the energy range
15--150~keV at $\approx 2$-week resolution, the \swi{}/XRT light curve 
in the energy range 2--10~keV, the \swi{}/UVOT light curve 
in the UV $W2$ filter, the \swi{}/UVOT light curve 
in the optical $B$ filter, the GASP-WEBT light curve 
in the optical $R$ filter, and the VLBI radio core
at 15~GHz (filled circles) and the UMRAO 14.5~GHz (open squares)
light curves, respectively.

We investigated the correlation between the \gray flux and the optical 
flux density in the $R$ band by means of the DCF method, 
computing the DCF on four distinct periods: July 2007 (mid 2007),
November--December 2007 (Fall 2007), May--August 2008 (mid 2008), and
October 2007 - January 2009 (Fall 2008). 
The period ``fall 2007'' offers a good opportunity to test the 
correlation, since the $\gamma$-ray flux, and even more the optical flux, 
exhibited strong variability, and the period of common monitoring lasted 
for more than a month. The corresponding DCF shows a 
maximum $\rm DCF \sim 0.38$ for a null time lag.
However, the shape of the peak is asymmetric, and a centroid
calculation yields a time lag of $-0.42$\,days. 
This result is in agreement with what was found by 
\cite{Donnarumma2009:3C454_dec07} and by \cite{Bonning2009:3c454_smarts}.

%%%%%%%%%%%%%%%%%%%%%%%%%%%
\section{Conclusions}   
%%%%%%%%%%%%%%%%%%%%%%%%%%%

Since July 2007, \source{} has been playing the same role for \agile{}
as 3C~279 had for EGRET, and during  the period July 2007 - January
2009 we acquired data not only in the \gray energy band, but across 14 
decades in energy. 
This allowed us to construct simultaneous SEDs, sampling high,
intermediate, and low \gray emission states, involving both ground and
space based observatories.

We found that the role of the external Compton on the disk and the
broad-line region radiation (and possibly also on hot corona 
photons) is crucial to account for the hard \gray spectrum states. 
Moreover, thanks to the extremely dense optical coverage provided by
the GASP-WEBT, we were able to study the correlations between the 
\gray and optical fluxes. We found a $\simeq 1$ day possible lag of the 
high energy photons with respect to the optical ones.

The simultaneous presence of two \gray satellites, \agile{} and Fermi, 
the extremely prompt response of wide-band satellites as \swi{}, 
and the long-term monitoring provided from the radio to the optical by 
the GASP-WEBT Consortium will assure the chance to investigate and
study the physical properties of \source{} and of several more 
blazars both at high and low emission states.

\acknowledgements %%% Text of acknowledgements runs on after this command.
I am grateful to L. Maraschi and to all the Como
Workshop SOC and LOC members for having organized such an excellent
and fruitful meeting. The results presented here were obtained
in collaboration with the AGILE AGN Working Group (A. Bulgarelli, 
A.W. Chen, F. D'Ammando, I. Donnarumma, A. Giuliani,
F. Longo, L. Pacciani, G. Pucella, V. Vittorini, and M. Tavani), 
the WEBT-GASP Team (in particular M. Villata and C.M. Raiteri), 
and the \swi{} Team (in particular P. Romano, H. Krimm, N. Gehrels, 
the duty scientists, and science planners).
The AGILE Mission is funded by the Italian Space Agency (ASI) with
scientific and programmatic participation by the Italian Institute
of Astrophysics (INAF) and the Italian Institute of Nuclear
Physics (INFN).
S.V. acknowledges financial support by the contracts ASI I/088/06/0
and ASI I/089/06/0.

%%% THE BIBLIOGRAPHY
%%%
%%% CONSULT SECTION 3 OF "INSTRUCTIONS FOR AUTHORS" FOR HOW TO USE NATBIB.
%%% AUTHORS ARE ENCOURAGED TO USE EITHER THE "THEBIBLIOGRAPY" ENVIRONMENT
%%% BY UNCOMMENTING (DELETING THE "%" SYMBOL) THE COMMANDS BELOW, OR BY
%%% USING THE BIBTEX ENVIRONMENT. TO FIND OUT WHICH IS APPLICABLE TO YOUR
%%% CONTRIBUTION, CONSULT THE VOLUME EDITORS FOR YOUR PROCEEDINGS.
%%%

\end{document}